# Room-temperature valley-selective emission in Si-$MoSe_2$ heterostructures enabled by high-quality-factor chiroptical cavities

Feng Pan[1,†,*], Xin Li[2,†], Amalya C. Johnson[1], Scott Dhuey[3], Ashley Saunders[4], Meng-Xia Hu[2], Jefferson P. Dixon[5], Sahil Dagli[1], Sze-Cheung Lau[6], Tingting Weng[7], Chih-Yi Chen[1], Jun-Hao Zeng[2], Rajas Apte[1], Tony F. Heinz[6,8], Fang Liu[4,*], Zi-Lan Deng[2,*], Jennifer A. Dionne[1,9,*]

[1] *Department of Materials Science and Engineering, Stanford University, Stanford, California 94305, United States*
[2] *Guangdong Provincial Key Laboratory of Optical Fiber Sensing and Communications, Institute of Photonics Technology, College of Physics & Optoelectronic Engineering, Jinan University, 510632, Guangzhou, China*
[3] *The Molecular Foundry, Lawrence Berkeley National Laboratory, Berkeley, California 94720, United States*
[4] *Department of Chemistry, Stanford University, Stanford, California 94305, United States*
[5] *Department of Mechanical Engineering, Stanford University, Stanford, California 94305, United States*
[6] *Department of Applied Physics, Stanford University, Stanford, California 94305, United States*
[7] *Marvell Technology, Inc., Santa Clara, California 95054, United States*
[8] *SLAC National Accelerator Laboratory, Menlo Park, California 94025, United States*
[9] *Department of Radiology, Stanford University, Stanford, California 94305, United States*

[†]These authors contributed equally

*Emails: fpan22@stanford.edu, fliu10@stanford.edu, zilandeng@jnu.edu.cn, jdionne@stanford.edu

## Abstract

Transition metal dichalcogenides possess valley pseudospin, enabling coupling between photon spin and electron spin for classical and quantum information processing. However, rapid valley-dephasing processes have impeded the development of scalable, high-performance valleytronic devices operating at room temperature. Here we demonstrate that a chiral resonant metasurface can enable room-temperature valley-selective emission in $MoSe_2$ monolayers independent of excitation polarization. This platform provides circular eigen-polarization states with a high quality factor (Q-factor) and strong chiral near-field enhancement. The fabricated Si chiral metasurfaces exhibit chiroptical resonances with Q-factors up to 450 at visible wavelengths. We reveal degrees of circular polarization (DOP) reaching a record high of 0.5 at room temperature. Our measurements show that the high DOP can be attributed to the significantly increased chiroptical local density of states, which enhances valley-specific radiative transition rates by a factor of ~13. Our work could facilitate the development of ultracompact chiral classical and quantum light sources.

# Introduction

Employing quantum mechanical two-level systems (qubits) for computation necessitates resolving the tension between two fundamental yet conflicting core requirements: long coherence times, and strong coupling to their environment for initialization and readout. Electronic degrees of freedom such as angular momentum (spin or orbital) have strong and tunable coupling, but coherence tends to be a limitation. In parallel, photon spin angular momentum (circular polarization) typically has a long coherence time but intrinsically weak coupling. If electrons and photons could be coupled together in materials, then it might be possible to achieve simultaneous strong coupling and coherence in two-level systems.

Monolayers of transition metal dichalcogenides (TMDCs) provide a promising platform to achieve such electron-photon coupling. The electrons in the valleys of their spin-orbit split bands can be selectively excited with circularly polarized light, resulting in valley polarized excitons with corresponding spin and valley indexes.[1,2,3] This new "valleytronic" route of encoding information offers an exciting approach for classical and quantum information storage, processing, and computation.[4,5] The spin-valley polarization of TMDC monolayers can be probed and evaluated using the degree of optical circular polarization (DOP), $\mathrm{DOP} = (I(\sigma^-) - I(\sigma^+))/(I(\sigma^-) + I(\sigma^+))$, where $I(\sigma^\pm)$ is the photoluminescence intensity recorded for left-handed or right-handed circularly polarized emission ($\sigma^\pm$ ).

In TMDCs, the prepared spin-valley polarization often undergoes rapid dephasing processes, including intervalley exchange coupling and phonon-assisted scattering, resulting in fast spin decoherence at room temperature.[6] Such dephasing prevents TMDCs from being employed as room-temperature on-chip valley-selective devices for applications in valleytronics[7] and quantum information systems (QIS).[8] Recently, it has been demonstrated that chiral light-matter interactions of TMDC monolayers can be enhanced by employing the near-fields of plasmonic and dielectric nanostructures. In some studies, the nanostructures are achiral, with the enhanced near field created with circular polarized excitation.[9–11] In other studies, the nanostructures are chiroptical, with broken in-plane or out-of-plane mirror symmetries.[12–17] Such chiral nanostructures enable initialization and readout of the TMDC valley polarization without relying on the optical excitation's helicity, and hence are especially promising for on-chip electro-optic transduction and quantum integrated photonics. The superchiral near-fields around nanostructures compete with intervalley scattering at the *K* or *K'* valleys in the TMDC, and thus enhance valley-selective emission.[12,13,15–18] However, in each of these chiral metasurface-aided studies, the intrinsic absorption loss and/or scattering loss of the plasmonic or Mie-resonant nanostructures limits the quality factor ($Q$-factor) of the resonance and the achievable chiral near-fields. Consequently, these TMDC-interfaced metasurfaces either have to operate at cryogenic temperatures to exhibit a meaningful DOP (e.g., DOP = 0.43 at T = 87 K)[13] or exhibit low DOP at room temperature (e.g., DOP < 0.10).[18]

Here, we design and fabricate Si nanostructures that break both in-plane mirror and inversion symmetries, resulting in high-$Q$ chiroptical resonances at visible wavelengths (**Figure 1a**). When interfaced with large-area $MoSe_2$ monolayers, we demonstrate strong valley-selective emission (i.e., a maximum DOP of 0.5 at room temperature independent of excitation polarization) through steady-state photoluminescence (sPL) measurements at elevated temperatures spanning 100 K up to room temperature. Our nanostructures are based on chiral quasi-bound states in the continuum (q-BICs), which localize light in the radiation continuum, resulting in both high $Q$-factor resonances[19] and maximum chirality.[20,21] Recent demonstrations of high-$Q$ planar and three-dimensional chiral metasurfaces show that chiral q-BIC modes can drive maximum circular dichroism (CD),[22–26] high-efficiency chiral light sources[22], optical modulators,[27] and nonlinear optics.[24] In our study, the chiral q-BIC localizes valley-selective emission in the vicinity of the photonic $\Gamma$-point, as demonstrated through reciprocal space mapping of the TMDC PL emission. Time-resolved PL (TRPL) measurements further reveal that the strong chiral near-field enhances the radiative recombination of excitons/trions at the *K'* valley over the *K* valley through a valley-selective Purcell effect. Compared with previous works based on either achiral individual[10] and extended cavities[9,17,28] or conventional chiral metasurface cavities,[13,16,18,29] our heterostructured platform enabled by chiral q-BICs exhibits both high *Q*-factors and record-high DOPs at room temperature (**Figure 1b**). Taken together, the high-*Q* chiroptical resonances and strong near-field valley-selective enhancement shown in our work are the key to achieving such superior performance, en-route to scalable high-performance valleytronic devices.

# Results

## Chiral q-BIC metasurfaces for valley-selective emission

To achieve photon-spin-valley locking in $MoSe_2$ monolayers, we designed a crystalline Si (c-Si) metasurface on a glass substrate. The metasurface is composed of periodically arranged nano-squares with etched corners across the anti-diagonal of the square (**Fig. 1a**). In contrast with conventional plasmonic and dielectric chiral metasurfaces,[12,13,16,18,30–35] q-BIC-driven chiral metasurfaces leverage symmetry-protected photonic states that exhibit, in principle, infinitely high *Q*-factors and strong helicity-selective near-field enhancement. The BIC state in our design is a vortex polarization singularity enclosed by elliptical eigen-polarizations[24] and consequently is uncoupled to far-field CPL. Tuning the asymmetry parameters breaks the in-plane inversion and mirror symmetries and enables the coupling of chiral BIC modes to free-space radiation, thus forming a chiral q-BIC resonance. All out-of-plane symmetries are preserved with an index-matching layer (i.e., polymethyl methacrylate (PMMA)) on top of the metasurface (**Fig. 1a**) to achieve the maximal planar chiroptical effects. Doing so creates a circular eigen-polarization state at the $\Gamma$-point surrounded by elliptical eigen-polarizations (**Supplementary Fig. 1**). **Figure 1c** shows the simulated transmission spectra of the Jones matrix elements on the circular polarization basis ($T_{ij}$, where $i = r, l$; $j = r, l$; $r$ represents right-handed circular polarization or RCP, $l$ represents left-handed circular polarization or LCP; $i$ and $j$ represent circular polarization states of outgoing and incoming light, respectively) for one of our designed c-Si metasurfaces. A high-*Q* chiroptical resonance at 772 nm is shown in the $T_{rl}$-component spectrum which is not seen in the rest of component spectra ($T_{lr}$, $T_{ll}$, and $T_{rr}$). Fitting $T_{rl}$ to the Fano expression (**Supplementary Fig. 2a**) yields a *Q*-factor of 677. Note that the transmission CD does not reach unity, mainly due to the interference between chiral q-BIC mode and background scattering channels (e.g., higher-order Mie modes). Multipole expansion of the chiral q-BIC suggests that the mode is mainly characterized by an electric quadrupole (EQ) (**Supplementary Fig. 3**).

In our design, the etched length and width differences in the *x* and *y* directions ($\Delta L = L1 - L2$, $\Delta W = W1 - W2$) are engineered to adjust *Q*-factors and chiroptical effects of the chiral q-BIC. The overall edge length ($L$) and height ($H$) of the nano-square determine the resonance wavelengths and ultimately the detuning between chiral q-BIC and valley exciton resonance in monolayer $MoSe_2$. **Figure 1d** reveals that smaller $\Delta W$, $\Delta L$ yield *Q*-factors exceeding 1,000, which would otherwise scale up rapidly in an inversely quadratic relationship with the asymmetry parameter (e.g., $\alpha = \Delta W/W1$) assuming zero nonradiative loss at visible wavelengths (**Supplementary Fig. 4**). To verify the role of chiral q-BIC mode in enabling valley-selective emission, an ensemble of in-plane emitting dipoles are employed to excite the chiral q-BIC mode (**Supplementary Note 1**). These dipoles are achiral and therefore populate *K* and *K'* valleys equally, but valley depolarization is ignored in these simulations. **Figure 1e** shows how our structure exhibits a nearly unitary emission CD, suggesting that the chiral q-BIC mode favors $\sigma^-$ emission over $\sigma^+$ emission in the metasurface's upper space, and just the opposite in the lower space (**Supplementary Fig. 5**). This behavior persists as we alter $\Delta W$ (**Supplementary Fig. 6**). **Figure 1f** shows the strong helicity-selective near-field enhancement (also see **Supplementary Fig. 7**, shown for a $\Delta W$ = 22 nm, $\Delta L$ = 2 nm, $L$ = 297 nm, $H$ = 220 nm). A striking contrast in the cross-sectional electric field enhancement is seen when the chiral q-BIC mode is excited (e.g., at 772 nm) by LCP versus RCP, confirming the role of the chiral q-BIC in differentiating the states of circular polarization. In order to evaluate the near-field chirality through the chiral metasurface structure, we computed optical chirality density (OCD) using $\mathrm{OCD} = -\frac{\omega}{2c^2} Im(\mathbf{E}^* \cdot \mathbf{H})$,[36] where **E** and **H** represent complex electric and magnetic fields, and $\omega$ is the angular frequency of light. Complementary to both unity emission CD and near-field chiroptical enhancement is the net near-field chirality identified on the plane just atop the metasurface structure where $MoSe_2$ monolayers reside (see **Figure 1g** and **1h**). Taken together, the chiral q-BIC mode in our design readily enables valley-selective emission in $MoSe_2$ monolayers.

We fabricated the c-Si chiral metasurfaces on a glass substrate. These c-Si thin films have much lower loss in visible and near-infrared frequencies than amorphous and polycrystalline Si.[37–39] To fabricate c-Si chiral metasurfaces on an isotropic substrate and gain optical access to the $MoSe_2$-integrated structures on both sides, we developed a wafer direct-bonding process to transfer c-Si thin films (thickness: 220 nm) from a silicon-on-insulator (SOI) wafer onto a double-sided polished borofloat glass wafer (**Methods** and **Supplementary Fig. 8**). The resulting c-Si films show much lower loss in the spectral region of interest (**Supplementary Fig. 9**), which is consistent with literature reports.[37,38] We then fabricated c-Si metasurfaces on the glass substrate using electron-beam lithography (**Methods**). All of the fabricated metasurfaces have a height of 220 nm defined by the bonded c-Si thin film. **Figure 1i** shows one of the fabricated metasurfaces in top-view and tilted-view scanning electron microscope (SEM) images, in which $P$ = 430 nm, $L$ = 297 nm, $\Delta L$ = 2 nm, $\Delta W$ = 22 nm. No noticeable difference between sharp-corner nanostructures (**Fig. 1a** and **1f**) and round-corner nanostructures due to uncertainty in patterning sharp edges (**Figure 1i**) is observed in simulated transmittance and electric field enhancement distribution **(Supplementary Fig. 10)**. We utilized a home-built optical setup (**Supplementary Fig. 11** and **Supplementary Note 2**) to measure the transmission spectra of the metasurfaces coated with PMMA. A high-$Q$ chiroptical resonance is seen at 772 nm with a Fano lineshape (green curve) as shown in **Figure 2a**, matching the $T_{\mathrm{rl}}$-component spectrum shown in **Fig. 1c**. Fitting the spectrum to the Fano expression (**Supplementary Fig. 2b**) yields a $Q$-factor of 304. Overall, the fabricated metasurfaces exhibit experimentally fitted $Q$-factors ranging from 200 to 450, and the resonances span the A-exciton emission wavelength range of $MoSe_2$ monolayers (**Supplementary Fig. 12**). These sets of high-$Q$ metasurfaces allow us to examine the influence of resonance detuning between chiral q-BIC mode and A-exciton upon valley-selective emission.

## Valley-selective emission at elevated temperatures

We exfoliated large-area $MoSe_2$ single-crystal monolayers from bulk crystals using the gold tape exfoliation method (**Methods**)[40] and subsequently transferred the monolayers onto our chiral metasurfaces and coated them with PMMA. An image for one of the $MoSe_2$-metasurface integrated devices is shown **Figure 2b,** with monolayers (pink and green) spanning both the metasurface array (yellow square) and glass substrate. The incomplete monolayer coverage on the metasurface is due to relatively weak adhesion between the transferred $MoSe_2$ monolayer and c-Si meta-atoms compared to that between the transferred monolayer and the relatively flat glass substrate. The integrated devices were characterized at elevated temperatures (100-294 K) using a home-built spin-resolved cryogenic PL setup (**Supplementary Fig. 13** and **Supplementary Note 2**). Central to assessing the valley-selective Purcell effect imposed solely by the metasurfaces is initializing exciton population equally at *K* and *K'* valleys. In our measurements, we excited the integrated devices with linearly polarized light at 633 nm (or 640 nm) and measured PL emission of $\sigma^+$ and $\sigma^-$ photons (see **Fig. 1a** and **Figure 2c**) through a combination of a quarter-wave plate and a Wollaston prism or a linear polarizer (**Methods**). We conducted hyperspectral PL imaging of the integrated devices, which allows us to assess the heterogeneity of the PL intensity, peak wavelength, and linewidth. These properties are associated with both monolayers' quality and the interplay between monolayers and meta-atoms.

By enforcing the chiral q-BIC mode resonant with valley-dependent exciton transitions in monolayer TMDCs, our chiral metasurface can enhance the $\sigma^-$ radiative transition rate from *K'* valley (**Fig. 2c**), regardless of the state of photoexcitation polarization. Similarly, $\sigma^+$ emission can be enhanced by redesigning the metasurfaces with the opposite chirality. **Figure 2d** shows the intensity maps of $\sigma^+$ and $\sigma^-$ emission obtained from sPL measurements, suggesting strong valley-selective emission. However, this behavior is not seen for bare monolayers on the glass or c-Si given very comparable intensity maps between $\sigma^+$ and $\sigma^-$ emission (see **Supplementary Fig. 14a** and **14b**). Please note that the PL intensity variation across the maps is primarily attributed to the incomplete monolayer coverage on the metasurface and the spectral variation across the metasurface array is negligible given a narrow spatial distribution of q-BIC resonance wavelengths (see **Supplementary Fig. 15**). To rule out any chiroptical effects near the excitation wavelengths that may contribute to chirally selective absorption, we simulated the transmission spectra of the Jones matrix elements and near-field electric field profiles. No chiroptical resonances are observed, and comparable electric field enhancements are obtained for both LCP and RCP excitation (**Supplementary Fig. 16** and **17**). Taken together, it is the chiral q-BIC mode at the excitonic emission wavelength that drives valley-selective emission. High-intensity pixels are correlated to the monolayer covered areas shown in **Fig. 2b**. Further characterization using SEM reveals that monolayer flakes are more inclined to be intact and consequently form interfacial domains (**Supplementary Fig. 18**).

**Figure 3a** displays sPL emission spectra for a single pixel (5 μm × 5 μm) obtained from photon-spin-resolved hyperspectral imaging of the $MoSe_2$-metasurface integrated device shown in **Fig. 2b** at different temperatures (100, 140, 200, and 294 K). We observe several changes in the PL with temperature, including red-shifted emission lineshapes, decreased intensity, and increased linewidth with increasing temperature, consistent with the literature reports on TMDC monolayers.[41,42] However, we also observe valley-selective emission behavior across all temperatures including, notably, room temperature. We also compare PL emission of $MoSe_2$ monolayers on the metasurfaces to that of bare monolayers on the glass substrate. The exciton/trion emission intensity is enhanced by a factor of close to 10 at temperatures ranging from 100 K to room temperature (see **Supplementary Fig. 19**), suggesting c-Si metasurfaces' Purcell enhancement. Each PL peak originates from an intrinsic Lorentzian lineshape broadened by a Gaussian distribution due to inhomogeneous broadening.[43] We fit the main exciton peak to a Voigt function, i.e. a convolution of a Lorentzian and a Gaussian,[44] and capture the shoulder feature (which is assigned to trion recombination) via a secondary Gaussian (see **Supplementary Fig. 20**). In order to quantify valley-selective emission, we use the single emission intensity values at peak wavelengths from the fits to compute maximum DOPs for individual pixels in the hyperspectral images. This analysis yields the spatial distribution of DOPs for the integrated devices at various temperatures (see **Figure 3b**). Unlike nearly zero DOPs seen for bare monolayers on the glass substrate (see **Supplementary Figs. 14c and 21**), high DOPs are obtained for monolayers on the metasurfaces. As the temperature increases, the overall DOP increases and reaches a record high of 0.5 at 294 K. This behavior is attributed to the resonance detuning between excitons and chiral q-BIC mode which is controlled by temperature-dependent exciton energy.

We have determined spatially averaged DOPs for all characterized integrated devices. **Figure 3c** shows spatially averaged DOP versus the resonance detuning ($\delta\lambda$) at elevated temperatures, where $\delta\lambda = \lambda_0 - \lambda_1$is the difference between the chiral q-BIC mode resonance wavelengths that we are able to measure only at 294 K ($\lambda_0$) using our optical setup (**Methods** and **Supplementary Fig. 11**) and the exciton resonance wavelengths at elevated temperatures ($\lambda_1$). For the integrated device presented in **Fig. 3b**, when the temperature increases, $\delta\lambda$ approaches zero and consequently the spatially averaged DOP reaches its maximum (see green stars in **Figure 3c**). The spatially averaged DOPs peak at a certain detuning when the integrated devices are measured at various temperatures. Only when measured at 294 K are the DOPs peaked near $\delta\lambda = 0$, which may be further boosted if the chiral q-BIC resonance is slightly tuned to the exciton resonance wavelength. At lower temperatures, the detuning at which maximum DOP is achieved shifts toward a more positive detuning. This detuning is attributed to the shifted chiral q-BIC mode resonance at different temperatures. The refractive index of Si decreases as the temperature decreases[45] and chiral q-BIC mode resonance wavelength is ultimately blue-shifted (see **Supplementary Fig. 22**). That being said, at a lower temperature, a maximum spatially averaged DOP is seen at a more positive detuning. The extent to which the chiral q-BIC mode resonance is blue-shifted observed in our simulations is less pronounced than maximum DOP's detuning shift determined in our experiments. The strain induced by PMMA's thermal response may contribute to this disparity, which is difficult to incorporate into simulations. These results highlight the importance of aligning the chiral q-BIC mode to exciton resonance for maximizing valley-selective emission. Additionally, mitigation of the unfavorable strain effect induced by PMMA's thermal response may explain the observed increase in spatially averaged DOPs with rising temperature. As a further test, we altered the state of excitation polarization (i.e., applying linear and circular polarizations) and examined the resulting valley-selective emission properties. We observe almost the same DOP spatial distribution and spatially averaged DOPs irrespective of the excitation polarization (**Supplementary Fig. 23**). In other words, the photon-spin-valley locking is robust against the way that valley exciton population is initialized.

Our metasurface design exhibits a circular eigen-polarization at the $\Gamma$-point surrounded by elliptical eigen-polarizations in the momentum space (**Supplementary Fig. 1**). To interrogate how the chiral q-BIC mode shapes radiation patterns from $MoSe_2$ monolayers, we conducted sPL measurements in momentum space and obtained angle-resolved emission of $\sigma^+$ and $\sigma^-$ photons from the integrated devices. PL emission from the $MoSe_2$ monolayer integrated on metasurfaces is localized in the vicinity of the $\Gamma$-point in addition to stronger $\sigma^-$ emission, as shown in **Figure 4a**. In contrast, the angular emission intensities for $\sigma^+$and $\sigma^-$ photons are almost the same across the momentum space for $MoSe_2$ monolayers on the glass substrate (**Figure 4b**). When the temperature increases from 140 K to 294 K, the PL is more localized in the momentum space despite its broadening along the wavelength axis (**Supplementary Fig. 24**). These observations are consistent with the above analysis and confirm the capability of the chiral q-BIC mode for the simultaneous control of valley-selective emission and radiation pattern.

## Valley-selective emission decay dynamics

We evaluated the radiative recombination rates from valley polarized excitons/trions at *K* and *K'* valleys in our integrated devices, in order to determine the valley-selective Purcell enhancement enabled by the chiral q-BIC mode. The experiments were carried out with photon-spin-resolved TRPL, which measures the decay profiles of $\sigma^+$ and $\sigma^-$ emissions. The total population relaxation dynamics is a sum of radiative and nonradiative recombination, respectively. **Figure 5a** shows PL decay profiles at elevated temperatures for a single pixel obtained from the spin-resolved hyperspectral imaging of the integrated device. As the temperature increases, PL emission of both $\sigma^+$ and $\sigma^-$ photons decay more rapidly. We fitted the TRPL data to a biexponential decay model, $A_1 e^{-t/\tau_1} + A_2 e^{-t/\tau_2}$, convoluted with the instrument response function (IRF), where $\tau_1$ and $\tau_2$ are the fast and slow decay time constants, respectively. $\tau_1$ captures PL decays for both excitons and trions, whereas $\tau_2$ is attributed to the recombination of defect-trapped excitons.[9] Our sPL measurements above demonstrate that PL emission of valley-polarized excitons and trions are strongly modified by the chiral q-BIC mode. Thus, we focus on how the chiral q-BIC mode alters $\tau_1$ (hereafter as $\tau$). Although the intrinsic radiative recombination rate for a bright exciton in TMDC monolayers lies in the range of a $ps^{-1}$, for our temperature range ($\geq$ 100 K), only a small fraction of the excitons in a thermal distribution lie in the light cone, leading to a much reduced effective radiative decay rate on the order of $ns^{-1}$ .[46,47] This longer radiative decay time leads to greater sensitivity to nonradiative decay channels, such as trapping of excitons at defects and, at higher densities, exciton-exciton annihilation processes.[46] The latter, however, is unlikely to be significant for our conditions of weak continuous-wave laser excitation.

We observe that $\sigma^-$ emission decays faster than $\sigma^+$ emission, with more pronounced contrast at elevated temperatures such at 294 K, as shown in **Figure 5a**. Overall, the emission lifetime $\tau$ for monolayers on metasurfaces is found to be significantly shorter than that for monolayers on the glass substrate (0.189 ns versus 0.398 ns at 294 K) (see **Supplementary Fig. 25**). Note that although the dominant decay channels are nonradiative, particularly at 294 K, the reduction of emission lifetimes is caused by enhanced radiative transition rates. The nonradiative decay processes are independent of chiroptical effects. We estimate that radiative decay rates for monolayers on metasurfaces increase by a factor of ~13 compared to that for monolayers on the glass substrate by assuming nonradiative decay is 10 times faster than radiative decay as reported by the literature.[48] Such strong valley-selective Purcell enhancement is enabled by high-$Q$ chiroptical resonances and strong chiral near-field enhancement. The spatial distribution of $\tau$ for $\sigma^+$ and $\sigma^-$ emission at different locations is displayed in **Figure 5b**. Despite its heterogeneity, the lifetime of $\sigma^-$ emission is shorter than that of $\sigma^+$ emission by a factor of up to nearly two at 294 K, indicating valley-selective decay dynamics at elevated temperatures. In contrast, there is no meaningful difference in the lifetimes for $\sigma^+$ and $\sigma^-$ emission from monolayers on the glass substrate (see **Supplementary Fig. 26**). Furthermore, to quantify valley-selective Purcell enhancement and valley-selective emission, we estimate the time-integrated valley polarization using the following phenomenological equation, $\rho = (\tau(\sigma^+) - \tau(\sigma^-))/(\tau_v + \tau(\sigma^+) + \tau(\sigma^-))$.[18] In this equation, $\tau_v$ is the intervalley scattering time (see **Supplementary Note 3** for additional details), which is < 10 ps at $\geq$ 100 K[18,44] and therefore is minimal compared to $\tau(\sigma^-)$ and $\tau(\sigma^+)$. The expression then reduces to $\rho = (\tau(\sigma^+) - \tau(\sigma^-))/(\tau(\sigma^+) + \tau(\sigma^-))$. **Figure 5c** shows the statistical distribution of spatially averaged $\rho$ versus $\delta\lambda$ identified in sPL measurements for all the integrated devices measured using TRPL. $\rho$ is temperature-dependent and shows a very similar trend as DOP versus $\delta\lambda$ in **Fig. 3c**. $\rho$ is also maximized at certain detuning when the integrated devices are measured at individual temperatures, additional evidence that chiroptical effects enhance valley-selective PL decay dynamics.

## Discussion

We have designed, fabricated, and characterized high-$Q$ chiral metasurfaces driven by chiral q-BIC modes to achieve strong room-temperature, valley-selective emission in $MoSe_2$ monolayers. This effect yields circularly polarized emission at room temperature and does not depend on the preparation of the system with valley polarized excitons, hence being present for excitation with arbitrarily polarized light. Both our sPL and TRPL measurements show that the high DOP spans a wide range of temperatures (from cryogenic temperatures to 294 K) and is maximized (a record-high DOP of 0.5 at room temperature) when the chiral q-BIC mode is aligned with the A-exciton resonance. The radiative recombination of excitons and trions is found to be more enhanced at one specific valley over the other. Our observations are attributed to the significantly increased chiroptical local density of states provided by the chiral q-BIC mode, which enhances valley-specific transition rates when coupled to valley excitons. The circular eigen-polarization state provided by the chiral q-BIC mode results in localized angular emission around the mode's $\Gamma$-point. Compared with previous reports,[9,10,12,13,15–18,28,29,49] our scalable, large-area heterostructured platform exhibits better performances spanning chiroptical resonances with record-high $Q$-factors (see **Supplementary Fig. 27**), a record-high DOP at room temperature, and polarization-independent operation (see **Supplementary Table S1**). Our structural design with maximal planar chirality and significantly high-$Q$ chiroptical resonances, in combination with simpler patterning and etching procedures, render our platform more advantageous in enhancing valley-specific light-matter interactions in monolayer TMDCs than other reported metasurfaces with three-dimensional chirality.[22,23,25] Our findings may facilitate the study of chiral electro-optic transduction, single-photon emission encoded with spin angular momentum of light, and spin-polarized chiral polaritons.[50] In the classical regime, this nanophotonic platform could enable chiral electroluminescence[51] and chiral light detection.[52] In parallel, if one can create quantum-confined valley excitons through electrostatic trapping[53,54] in the vicinity of the chiral q-BIC mode, scalable on-demand chiral quantum light sources might be obtained. The valley-selective light-matter coupling, if further increased perhaps by engineering the light-matter interaction strength (e.g. using high-$Q$ silicon nitride cavities or a helicity-preserving Fabry-Pérot microcavity[55,56]), could result in hybridized states between valley-specific transitions and the chiral q-BIC mode, i.e. chiral polaritonic states, which may ultimately lift the energy degeneracy between $K$ and $K'$ valleys for significantly mitigating intervalley scattering and lead to nearly unity DOP.

## Methods

### Sample fabrication

Silicon-on-insulator and borofloat glass wafers (4") were first cleaned in a piranha solution. The wafers' surfaces were activated through $N_2/O_2$ reactive ion etching before they were directly bonded using a wafer bonder (Electronics Vision 501) at 400 °C under a pressure of 2000 N and a high vacuum ($10^{-6}$ torr). The bonded wafers were processed via a wafer grinder (Disco Backgrinder DAG810) to remove the majority of Si handling layer and finished by $XeF_2$ etching (Xactix e-1) to leave the thermal oxide on thin-film Si layer. The thermal oxide was removed by 6:1 BOE etchant to yield 220-nm Si-on-glass wafers. Diced chips from the Si-on-glass wafers were exposed using electron beam lithography (Raith EBPG5200). Specifically, an electron-beam resist, hydrogen silsesquioxane (HSQ), was spin-coated on 220-nm Si-on-glass chips and baked at 90 °C for 2 minutes. We coated a layer of electron conductive layer (espacer) for exposure before being developed in a salty develop solution (1% NaOH:4% NaCl). The sample was then etched by an inductively coupled plasma of $Cl_2$/HBr and then HBr (Oxford III-V etcher) and the resist was removed in a 2% HF solution afterwards.

The monolayer $MoSe_2$ is prepared with a gold tape exfoliation technique.[40] In brief, a 100 nm thick flat and clean gold layer is deposited on a clean Si wafer, which is later used as the gold tape. The gold tape is covered with a layer of polyvinylpyrrolidone (PVP) for protection against contaminations and a thermal release tape as a handling layer. A flat and clean gold surface is obtained from peeling the TRT/PVP/Au layer off the Si substrate. The freshly prepared gold surface is brought in contact with the $MoSe_2$ bulk crystal (HQ graphene) and exfoliates off a monolayer. The stack is transferred onto a $SiO_2$/Si chip followed by heating to remove the thermal release tape. The PVP layer is dissolved with water, and the gold layer is dissolved with $KI/I_2$ etchant solution. After preparation, the monolayers were then transferred to metasurfaces with cellulose acetate butyrate (CAB) using a wedging transfer technique.[57] Polymethyl methacrylate (PMMA) was coated before and after transferring $MoSe_2$ monolayers for transmission characterization and cryogenic PL measurements, respectively.

Numerical simulations

The transmission spectra of the Jones matrix elements and momentum-space eigen-polarizations were simulated with the finite element method (COMSOL Multiphysics Wave Optics module). c-Si thin films and borofloat glass substrate were characterized using an ellipsometer (J. A. Woollam) to obtain optical constants ($n$ and $k$) which were used throughout the simulations. The value for PMMA is set as $n$ = 1.485. Multipole expansion analysis was performed with the FDTD method (Lumerical).

Transmission characterization of fabricated metasurfaces

A supercontinuum laser (NKT Photonics Extreme Supercontinuum) delivered a Vis/NIR broadband light source (500-1100 nm) through an output of a spectral splitter (SuperK Split). The light source first passed through a pinhole to achieve a 1-mm beam size and then a set of linear polarizers (Union Optic PGT5012) and quarter-wave plate (Union Optic WSA4220-380-1100-M25.4) to generate circularly polarized light. The beam was weakly focused by a low-NA objective (Olympus 10x NA 0.25) onto a sample to achieve a beam size of ~80 um. The effective NA is significantly less than 0.25 to eliminate spin-orbit coupling and excitation of higher-order modes at high k-vectors (see **Supplementary Note 2** for more details). The sample was mounted on a motorized rotation stage. The transmitted light was captured by a second low-NA objective and passed through a polarization analyzer composed of another set of linear polarizer and quarter-wave plate. The transmitted light was focused onto the entrance slit of a spectrophotometer (Princeton Instruments IsoPlane 160) and imaged by a Si CCD camera (Princeton Instruments PIXIS 400BRX).

Steady-state photoluminescence characterization

In our steady-state photoluminescence (PL) measurements a continuous-wave (cw) laser beam (CNI MDL-III-633L) at 633 nm first propagated through a half-wave plate (Newport 10RP42-2), a linear polarizer (Union Optic PGT5012), and a quarter-wave plate (Union Optic WSA4220-380-1100-M25.4) to generate linearly or circularly polarized light. The beam was then focused to a diffraction-limited spot (see **Supplementary Note 2** for more details) by overfilling the back aperture of a high-NA objective (Olympus

MPLFLN50x NA 0.8) installed in CryoAdvance 100 (Montana Instruments). The excitation power was measured to be 5 μW prior to entering the objective and kept constant throughout all of the measurements. PL was collected through the same objective and passed through a 50:50 beam splitter, a notch filter (Edmund Optics 632.8 nm) with a transmission wavelength range of 475-845 nm, a long-pass filter (Edmund Optics 725 nm), and then a quarter-wave plate (Union Optic WSA4220-380-1100-M25.4). Circularly polarized PL emissions ($\sigma^+$ and $\sigma^-$) were converted to *s*- and *p*-polarized light which co-propagated through two different doublet achromatic lenses ($f_1$ = 200 mm and $f_2$ = 150 mm) before passing a Wollaston prism (Thorlabs WPQ10). The $\sigma^+$ and $\sigma^-$ emissions were then spatially separated. For real-space PL spectroscopy, spatially separated emissions were focused through a third doublet achromatic lens ($f_3$ = 100 mm) onto the entrance slit of a spectrophotometer (Princeton Instruments Acton SP2500) and then detected by a liquid-$N_2$ cooled Si CCD camera (Princeton Instruments PyLon 400B eXcelon). Thus, $\sigma^+$ and $\sigma^-$ light emissions were measured using a single camera for balanced detection. Raster-scanning a sample using the Rook nanopositioner stack while acquiring PL emission enables hyperspectral imaging of the integrated devices. For Fourier-space spectroscopy, spatially separated emissions entered the entrance slit of the spectrophotometer which was positioned at the back focal plane of the second lens.

Time-resolved photoluminescence characterization

A fiber-coupled pulsed laser (PicoQuant Prima) at 640 nm (pulse width 100 ps) was co-aligned with the cw 633-nm laser and propagated through the same optical components for excitation. The average excitation power was maintained at 0.21 μW which translated to an intensity of 8.3 μJ/cm$^2$ on the sample, which was employed throughout all of the TRPL measurements. The pulse repetition rate was 20 MHz. The instrument response function (IRF) was obtained by measuring laser light reflected on the sample. The emissions propagated through the same beam splitter and quarter-wave plate before entering a second channel for TRPL measurements. A linear polarizer (Union Optic PGT5012) mounted on a motorized rotation stage (Thorlabs PRM1Z8) was controlled for selecting s- or p-polarized light emissions to impinge onto a fiber launching system and then detected by a fiber-coupled single photon avalanche photodiode (Excelitas SPCM AQRH14 SPAD). PL decay profiles were obtained by tagging the arrival time of each PL photon to the laser trigger pulse using a time-correlated single photon counting module (PicoHarp 300) with a time resolution of 4 ps. The data were re-binned by a factor of 4 which equals a temporal resolution of 16 ps.

## Data Availability

Data used to produce the plots within this paper is available at https://zenodo.org/records/17196344. All other data used in this study are available from the corresponding authors upon request.

## Code Availability

The code used to produce the plots within this paper has been deposited in Zenodo (https://zenodo.org/records/17196344). Further code is available from the authors upon request.

## Acknowledgements

F.P. and J.A.D. acknowledge the support from the US Department of Energy, Office of Basic Energy Sciences (award DE-SC0021984) for computational design, the part of device fabrications, and optical measurements. S.D., S.-C.L., C.C., and T.F.H. acknowledge the support from the Office of Naval Research under the Multi-University Research Initiative (MURI) program (award N00014-23-1-2567) for optical measurements and data analysis. The construction of optical characterization setups was supported by the US Department of Energy, Office of Science, National Quantum Information Science Research Centers. S.D. was also supported by the US Department of Defense through the National Defense Science and Engineering (NDSEG) Fellowship Program. Work was performed in part at the Stanford Nanofabrication Facility (SNF) and the Stanford Nano Shared Facilities (SNSF), supported by the National Science Foundation (awards ECCS-1542152 and ECCS-2026822, respectively). Work at the Molecular Foundry was supported by the Office of Science, Office of Basic Energy Sciences, of the U.S. Department of Energy under Contract No. DE-AC02-05CH11231. X.L., M.H., J.Z. and Z.D. acknowledge the support from the National Natural Science Foundation of China (award 62422506, 12474383), Guangdong Basic and Applied Basic Research Foundation (award 2022B 1515020004), Guangdong Provincial Quantum Science

Strategic Initiative (No. GDZX2406004), and Guangzhou Science and Technology Program (award 2024A03J0465, 2025A04J5776). A.C.J., A.S., and F.L. acknowledge the support from the Defense Advanced Research Projects Agency (DARPA) under Contract No. HR00112390108 for the preparation of monolayer materials and construction of 2D layered structures. We also acknowledge helpful discussions with Chris Anderson and Harald Giessen.

## Author Contributions Statement

F.P. and X.L. contribute equally to this work. F.P. and J.A.D. conceived the study. X.L., F.P., M.H., J.P.D. and J.Z. conducted the theory and numerical simulations under the supervision of Z.-L.D. and J.A.D. F.P. and S. Dhuey fabricated the devices with the help from S. Dagli. A.C.J. and A.S. prepared and transferred the monolayers under the supervision of F.L. F.P. built up optical characterization setups and conducted the experiment with the help from S.L. and R.A. under the supervision of F.L., J.A.D. and T.F.H. F.P. analyzed the data with the help from T.W. and C.C. under the guidance of F.L. and J.A.D. F.P. wrote the manuscript with crucial input from all authors.

## Competing Interests Statement

The authors declare no competing interests.

## Additional information

Supplementary Information is available for this paper. Correspondence and requests for materials should be addressed to F.P. (fpan22@stanford.edu), F.L. (fliu10@stanford.edu), Z.-L.D. (zilandeng@jnu.edu.cn), and J.A.D. (jdionne@stanford.edu).

**Figure 1. Chiral q-BIC metasurface designed for room-temperature valley-selective emission**. (a) Schematic of the integrated Si-$MoSe_2$ device for controlling spin-photon coupling and generating valley-selective emission (red arrow) at the exciton resonance wavelength of $MoSe_2$ under arbitrarily polarized light excitation at 633 nm or 640 nm (green arrow). The periodicity ($P$) of the chiral q-BIC metasurface is 430 nm. PMMA: polymethyl methacrylate. (b) DOPs and $Q$-factors of some typical heterostructured TMDC-metasurface studies compared to our work. DOP: degree of optical circular polarization. R.T.: room temperature. Solid squares: achiral metasurfaces; Solid circles: chiral metasurfaces; Blue: low-temperature operation; Red: room-temperature operation. (c) Simulated transmission spectra of the Jones matrix elements ($T_{rl}$, $T_{lr}$, $T_{ll}$, and $T_{rr}$) given the parameters presented in **d**, showing a high-$Q$ (677) chiroptical resonance at 772 nm. $r$ represents right-handed circular polarization; $l$ represents left-handed circular polarization. $T_{rl}$: green solid line; $T_{lr}$: red solid line; $T_{ll}$: blue-green solid line; $T_{rr}$: black dashed line. (d) Simulated $Q$-factors versus asymmetry parameter ($\alpha$). The power-law fit (gray solid line) yields $Q \propto 1/\alpha^{0.26}$. (e) Simulated emission spectra of an ensemble of randomly located and oriented achiral dipoles (number: 200) coupled to the chiral q-BIC mode. Note that the spectra are acquired in the metasurface's upper space. $\sigma^{\pm}$: left-handed (blue solid line) or right-handed (red solid line) circularly polarized emission. (f) Electric field enhancement distribution for the design of a notched nano-square within a unit cell under LCP (left) and RCP (right) excitation. $H$ = 220 nm, $L1$ = 130 nm, $L2$ = 128 nm, $W1$ = 130 nm, $W2$ = 152 nm, $L$ = 297 nm. LCP: left-handed circular polarization; RCP: right-handed circular polarization. (g) Cross-sectional distribution of normalized OCD atop the chiral metasurface structure at $Z$ = 110 nm. OCD: optical chirality density. Note that the $MoSe_2$ monolayers are in contact only with the top surface of the structure at $Z$ = 110 nm, rather than occupying the entire internal volume. The origin of the coordinate is at the center of the nanostructure. (h) Surface-integrated OCD over the *xy*-plane along the *z*-axis. The location where $MoSe_2$ monolayer resides shows a positive surface-integrated OCD. (i) Top-view (left) and tilted-view (right) scanning electron microscope images of a chiral metasurface. Scale bar: 500 nm.

**Figure 2**. **Experimental characterizations of fabricated chiral q-BIC metasurfaces and integrated devices.** (a) Experimentally measured transmittance ($T$) spectrum for the metasurface shown in **Fig. 1c** and PL emission spectrum for A-exciton of monolayer $MoSe_2$ in an integrated device. The PL emission spectrum is normalized to its peak intensity. PL: photoluminescence. $\lambda$: wavelength. (b) An optical image of a chiral metasurface array (yellow) integrated with $MoSe_2$ monolayer flakes (pink areas). The yellow voids are holes in the monolayer. Scale bar: 50 μm. (c) Valley-selective emission for monolayer $MoSe_2$ coupled to the chiral metasurface under linearly polarized photoexcitation. $\sigma^{\pm}$: left-handed or right-handed circularly polarized emission. Only conduction and valence bands involving A-exciton radiative transitions are shown for the scope of this work with the spin-splitting of the bands omitted. (d) PL intensity ($I$) mapping for the integrated device at 294 K with a field of view defined by a black square shown in (**b**). The photoexcitation is at 633 nm. Scale bar: 25 μm.

**Figure 3**. **Valley-selective steady-state PL emission at elevated temperatures.** (a) PL spectra (dots) and fits (solid lines) for $\sigma^{+}$ (blue) and $\sigma^{-}$ (red) emission obtained from the photon-spin-resolved hyperspectral imaging at elevated temperatures with the same excitation power (5 μW) at 633 nm. $\lambda$: wavelength. (b) Spatial images of DOP for the integrated device at elevated temperatures. DOP: degree of optical circular polarization. The green circles indicate the location from which the PL spectra in **a** were collected. Scale bar: 20 μm. (c) Statistical distribution of spatially averaged DOPs over all the pixels in **3b** versus the detuning ($\delta\lambda$) between chiral q-BIC mode and exciton resonances. The horizontal error bars are the standard deviations of individual detuning measurements. The shaded areas are the standard deviations of spatially averaged DOPs at individual values. The green stars represent DOPs for the selected integrated structure shown in **b**.

**Figure 4**. **Angle-resolved valley-selective emission.** Momentum-resolved PL emission spectra for monolayers on the metasurface (**a**) and on the glass substrate (**b**) at 140 K. Emission intensity (I) is normalized to the maximum. The emission bands at 820-840 nm are attributed to defect-bound excitons. $\lambda$: wavelength; deg: degree. $\sigma^{\pm}$: left-handed or right-handed circularly polarized emission.

**Figure 5**. **Valley-selective PL decay dynamics at elevated temperatures.** (a) PL decay profiles (blue dots: $\sigma^{+}$; red dots: $\sigma^{-}$) and biexponential fits (blue lines: $\sigma^{+}$ fit; red lines: $\sigma^{-}$ fits) convoluted with the IRF (black dashed lines) for $\sigma^{+}$ and $\sigma^{-}$ emission obtained at elevated temperatures with the same excitation intensity. IRF: instrument response function. T: time. (b) Spatial images of $\tau$ (ns) for the integrated device. Scale bar: 20 μm. The green circles indicate the location from which the PL spectra in **a** were collected. Note that the relatively less spatial heterogeneity is due to narrower distributions of $\tau$ unlike the spatial images of DOPs shown in **Figure 3b**. $\sigma^{\pm}$: left-handed or right-handed circularly polarized emission. $\tau$: emission lifetime. (c) Statistical analysis of time-integrated valley polarization ($\rho$) versus the detuning ($\delta\lambda$, obtained from **Figure 3c**) between chiral q-BIC resonance and exciton resonance. The horizontal error bars are the standard deviations of individual detuning measurements. The shaded areas are the standard deviations of spatially averaged $\rho$ at individual $\delta\lambda$ values. The green stars represent $\rho$ for the selected integrated structure shown in **b**.

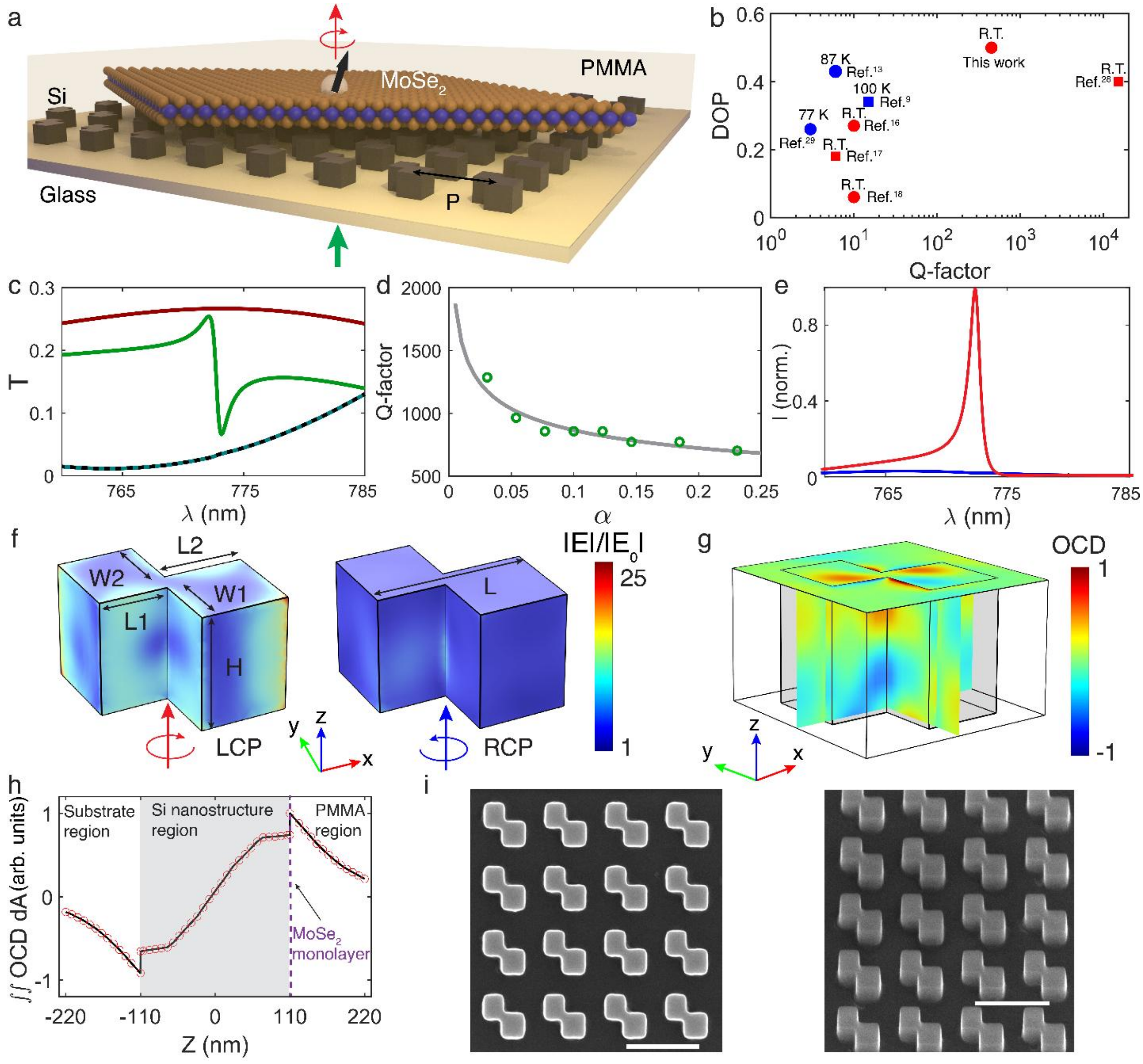

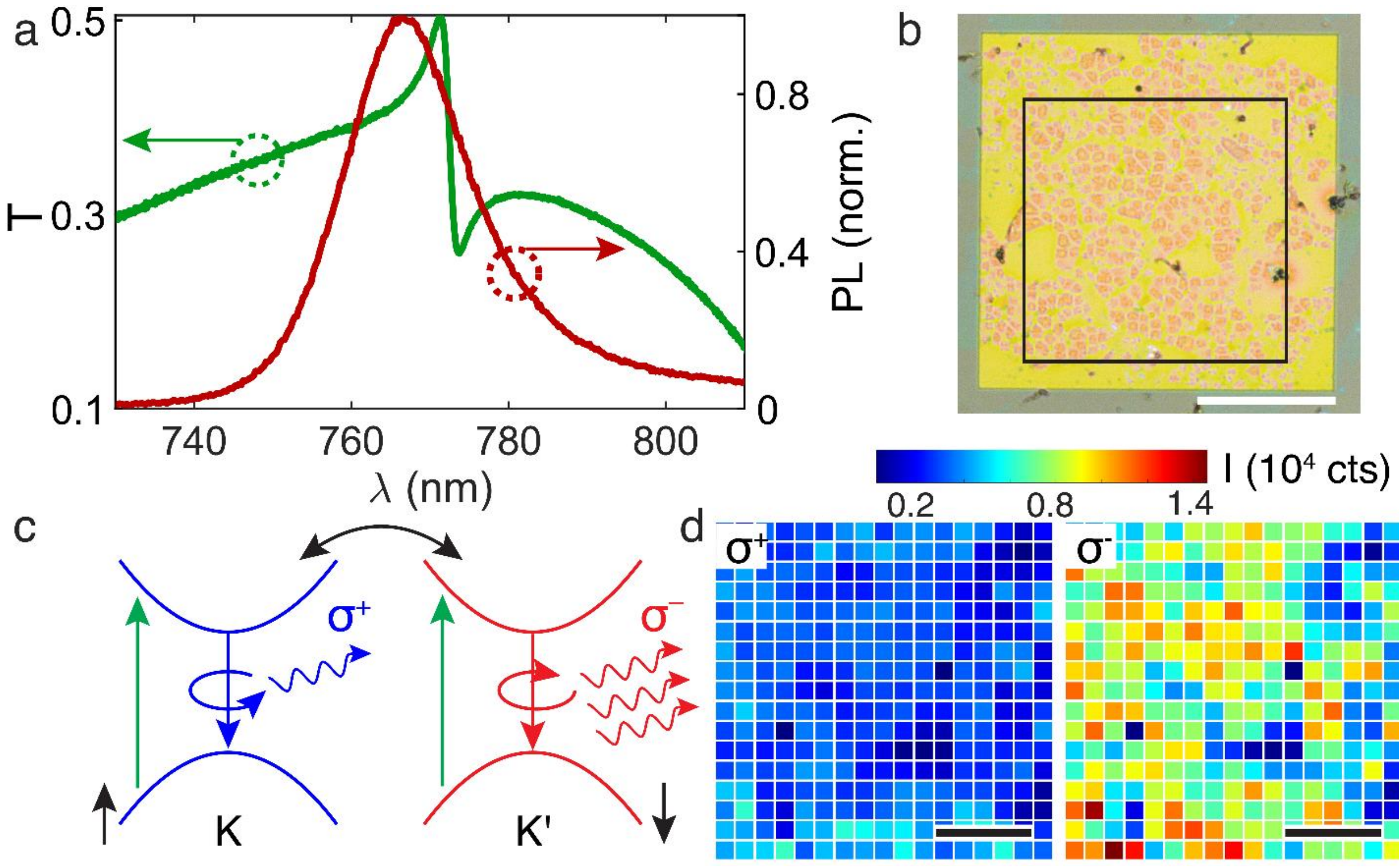

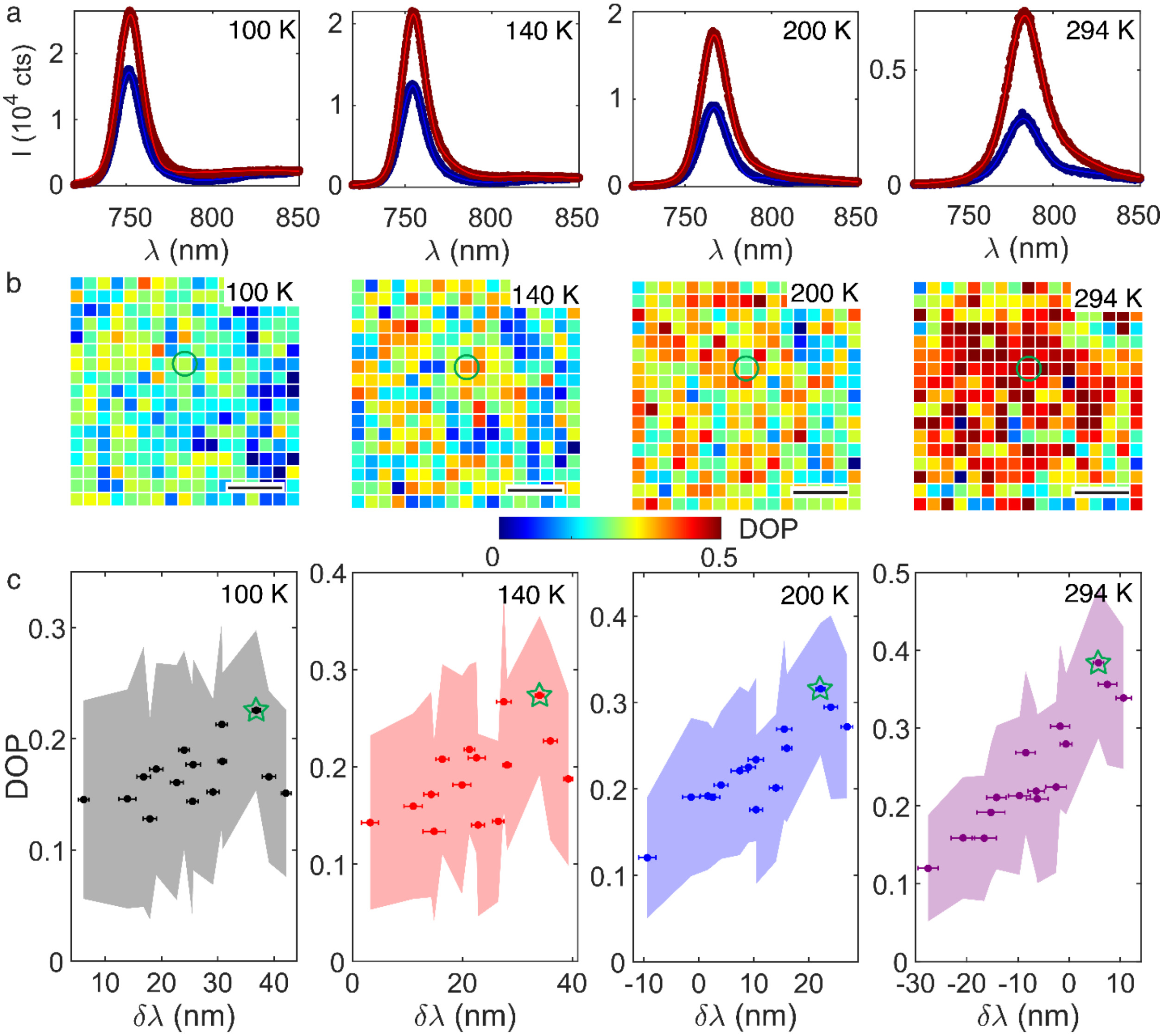

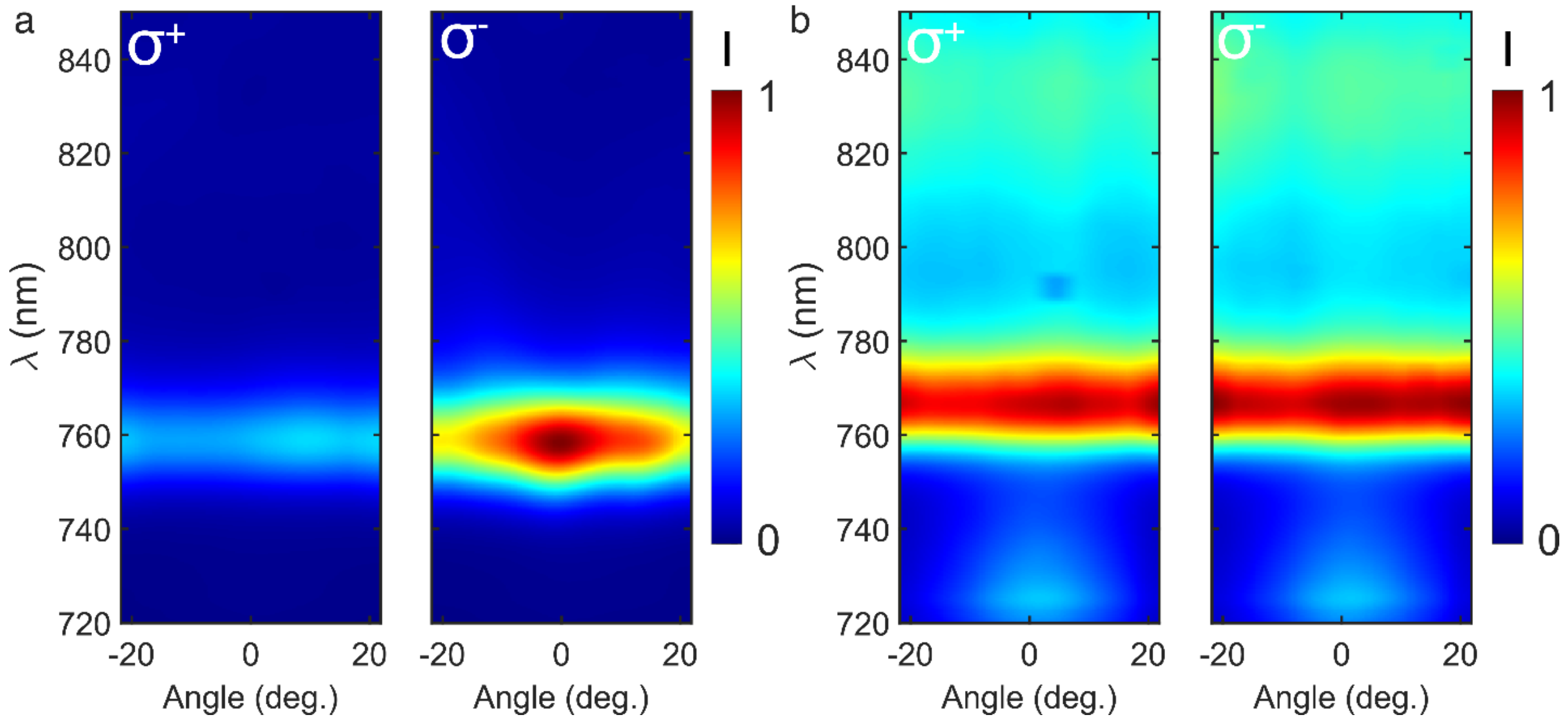

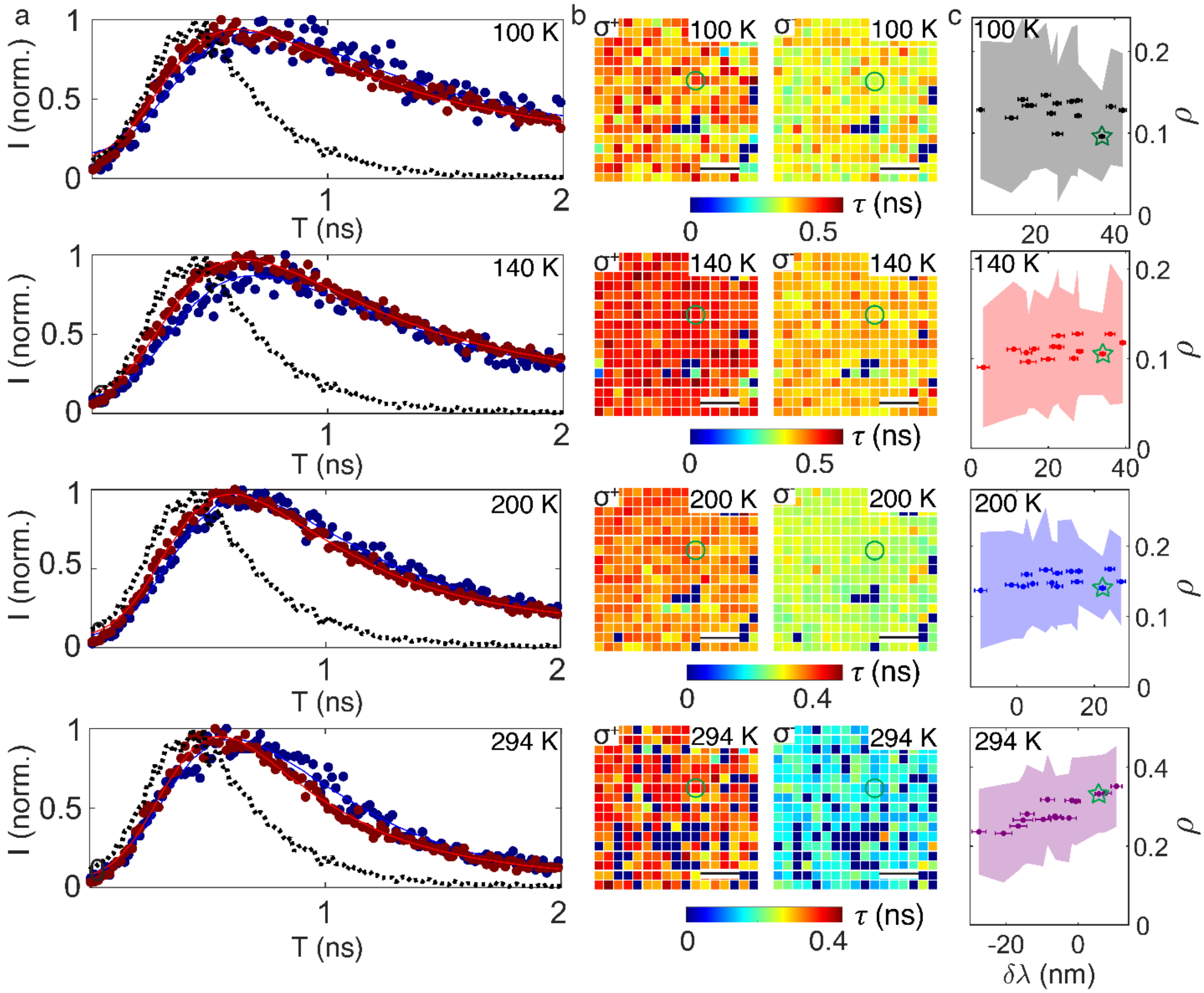